\documentclass[sigconf, nonacm, anoymous]{acmart}
\usepackage{graphicx}
\usepackage{multirow} 
\usepackage{rotating} %
\usepackage{float} 
\usepackage[inkscapelatex=false]{svg}
\usepackage{enumitem}

\AtBeginDocument{%
  }

\setcopyright{acmlicensed}
\copyrightyear{2018}
\acmYear{2018}
\acmDOI{XXXXXXX.XXXXXXX}

\acmConference[Conference acronym 'XX]{Make sure to enter the correct
  conference title from your rights confirmation emai}{June 03--05,
  2018}{Woodstock, NY}
\acmISBN{978-1-4503-XXXX-X/18/06}

\begin{document}

\title{ 
Intelligent Tutors for Adult Learners: An Analysis of Needs and Challenges }

\author{Adit Gupta}
\affiliation{%
  \institution{Drexel University}
  \streetaddress{3230 Market Street}
  \city{Philadelphia} 
  \state{PA}
  \country{USA}}

\author{Momin Siddiqui}
\affiliation{%
  \institution{Georgia Institute of Technology}
  \streetaddress{North Avenue}
  \city{Atlanta}
  \state{GA}
  \country{USA}}

\author{Glen Smith}
\affiliation{%
  \institution{Georgia Institute of Technology}
  \streetaddress{North Avenue}
  \city{Atlanta}
  \state{GA}
  \country{USA}}

\author{Jennifer Reddig}
\affiliation{%
  \institution{Georgia Institute of Technology}
  \streetaddress{North Avenue}
  \city{Atlanta}
  \state{GA}
  \country{USA}}

\author{Christopher MacLellan}
\affiliation{%
  \institution{Georgia Institute of Technology}
  \streetaddress{North Avenue}
  \city{Atlanta}
  \state{GA}
  \country{USA}}

\begin{abstract}

This work examines the sociotechnical factors that influence the adoption and usage of intelligent tutoring systems in self-directed learning contexts, focusing specifically on adult learners. The study is divided into two parts. First, we present Apprentice Tutors, a novel intelligent tutoring system designed to address the unique needs of adult learners. The platform includes adaptive problem selection, real-time feedback, and visual dashboards to support learning in college algebra topics. Second, we investigate the specific needs and experiences of adult users through a deployment study and a series of focus groups. Using thematic analysis, we identify key challenges and opportunities to improve tutor design and adoption. Based on these findings, we offer actionable design recommendations to help developers create intelligent tutoring systems that better align with the motivations and learning preferences of adult learners. This work contributes to a wider understanding of how to improve educational technologies to support lifelong learning and professional development.

\end{abstract}


\maketitle

\section{Introduction}
Intelligent tutoring systems support learners by guiding and scaffolding problem solving across a variety of domains \cite{maclellan2022domain}. 
Tutors have a well-documented history of improving student learning, particularly in K-12 settings \cite{Pane2014}, given to their ability to provide personalized instruction, immediate feedback, and adaptive support. In theory, these benefits should extend to adult learners in non-traditional educational contexts --- such as workplace training, online education platforms, and continuing professional development. However, tutors remain largely underutilized in these settings, either due to a lack of adoption or a limited body of research on their effectiveness for adult learners \cite{bernacki2021systematic}. Our work seeks to address this gap by exploring how tutors can support learning in adult education.

While the term {\it adult learner} colloquially refers to students over the age of 18, we define it here as individuals who are both over 18 and classified as non-traditional students---those pursuing education outside the conventional schooling pathway (e.g., a four-year undergraduate degree immediately after high school). This definition includes students who enter the workforce after high school and then return to college to change careers (e.g., through a two-year technical degree). Unlike traditional students, adult learners often juggle multiple responsibilities, such as full-time employment or family commitments.
Recent research on the adoption of educational technologies by adult learners indicates that this population remains understudied \cite{bernacki2021systematic}. 

Unlike K-12 students, who are often required to use tutors as part of their curriculum, adult learners self-regulate their own learning and are more strongly driven by intrinsic motivation \cite{lyndgaard2024towards}. This highlights the importance of investigating how tutors are utilized in voluntary, non-mandated settings, where engagement stems from personal goals rather than external obligations.  

While prior work on tutor adoption across a variety of user segments hypothesizes that tutors have not been widely adopted due to their high cost of development \cite{maclellan2022domain,murray2003}, this work explores an alternative hypothesis: that tutors are not adopted due to sociotechnical challenges that arise from a mismatch between the perceptions and needs of learners and the capabilities of intelligent tutors. Accordingly, our work is guided by two research questions: 

\begin{enumerate}
    \item[{\bf RQ1:}] Will adult learners use intelligent tutors when provided as a supplementary learning resource? 
    \item[{\bf RQ2:}] What are the specific needs of adult learners when using educational tools?
\end{enumerate}

To explore both questions, we developed an intelligent tutoring system based on best practices, called Apprentice Tutors. We then conducted a two-part study. The first part introduces Apprentice Tutors and examines whether students adopt the system. The second part evaluates whether the tutor meets users' needs and explores what insights can be gained about adult learning needs from its deployment. To investigate the unique needs of users, we conducted focus groups with both students and teachers. Following the focus groups, we performed a thematic analysis of the data \cite{Braun2006} and synthesized key findings into recommendations to improve the adoption and effectiveness of intelligent tutors in adult learning. The aim of this work is to guide designers and developers in building intelligent tutors with a better understand of needs of end-users and building pedagogical tools that allow for greater overall adoption and impact.

\section{Related Works}
Intelligent tutors are praised for their ability to support personalized learning at scale. Although human tutoring is particularly effective, it is cost prohibitive and therefore not available to everyone. Some researchers argue that private human tutoring can disproportionately benefit those with the financial means to afford such services, thereby exacerbating existing educational inequities \cite{BrayLykins2012,Bray2013Shadow,TanselBircan2006,ByunPark2012}. There have been several studies where intelligent tutors have shown success when used by K-12 learners in controlled settings where all participants engage in a fixed amount of activity \cite{koedinger1997intelligent,heffernan2014assistments,aleven2006cognitive, assistments_paper, oa_tutor}. 

For example, a randomized trial involving an algebra tutor was conducted as part of the Pittsburgh Urban Math Project. The study found that, on average, 470 students in the tutored group outperformed students in the non-tutored group by 15\% on standardized tests and achieved a 100\% improvement in the concepts targeted by the tutoring system \cite{koedinger1997intelligent}. Similarly, a large-scale randomized controlled trial of Carnegie Learning's Cognitive Tutor Algebra I found significant gains in student achievement, with tutored students outperforming their peers in traditional classrooms \cite{Pane2014}.

While certain pedagogical principles remain consistent across age groups, such as the benefits of high-quality feedback \cite{hattie2007power, shute2008focus, sybley_tutor}, it is important to recognize that adult learners differ from K-12 learners in ways that affect learning.
The context of adult learners diverges substantially.
Often, they have to balance education with many other life factors. Further, adult learners are much more self-directed and intrinsically motivated than K-12 learners \cite{kanfer2022learning}.
From a cognitive development perspective, some abilities, such as memory and abstract reasoning, tend to decline after early adulthood, while others, such as crystallized knowledge, continue to increase through midlife \cite{baltes1997lifespan, schaie2005developmental}. These developmental changes require considerations for age-inclusive instruction. Considering both the cognitive changes of the lifespan and the contextual differences in learning environments, it becomes evident that the research and practices developed for tutors may not be fully transferrable to adult learners. Although organizational researchers have provided guidelines for instructional design in technology-supported training contexts \cite{salas2012science}, it remains unclear how or to what extent these principles have been applied to develop effective tutors for adults. This knowledge gap highlights the need for further research into the specific requirements and considerations for designing tutors that meet the unique needs and characteristics of adult learners. 


\begin{figure*}
    \centering
    \includegraphics[width=1\textwidth]{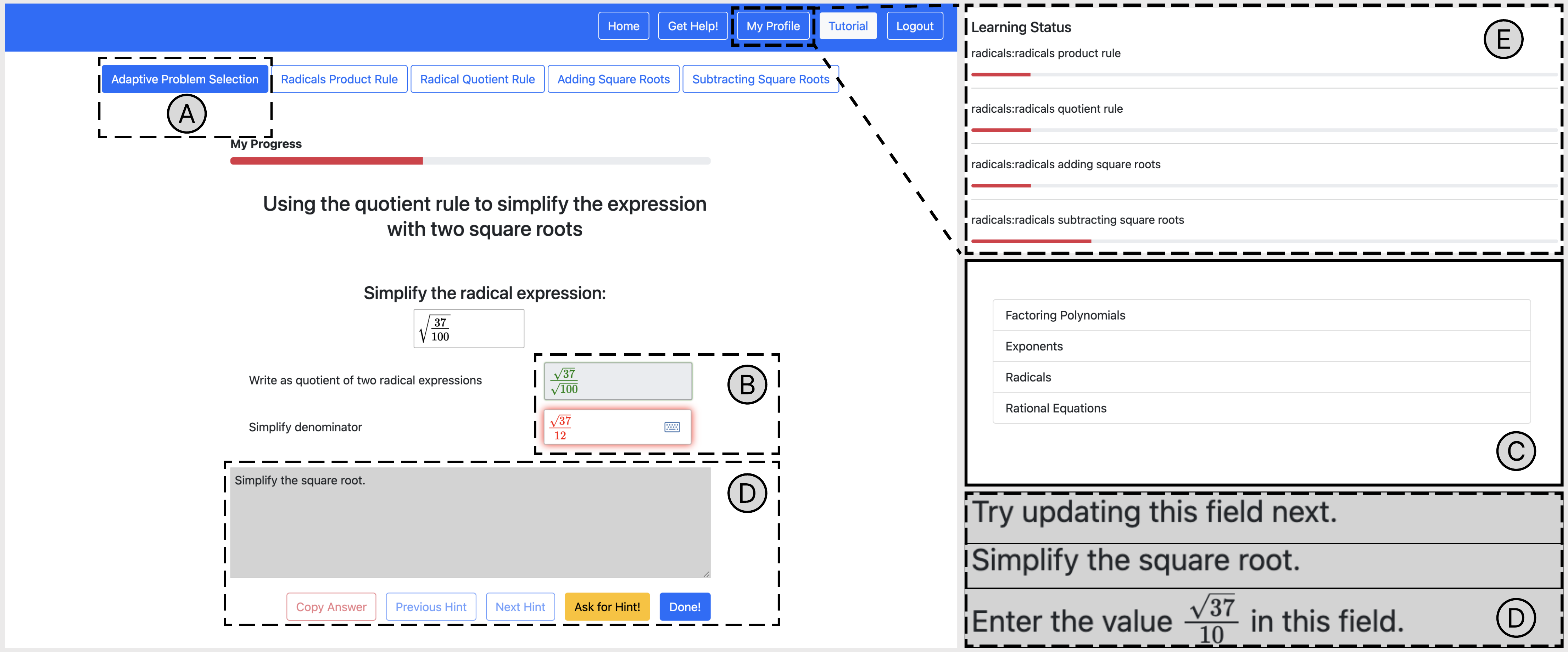} 
    \caption{User interface of Apprentice Tutors platform with key features: (a) penalization through adaptive problem selection (b) real-time correctness feedback (c) four available tutors (d) hint box and multi-layer hints (e) user profile screen with progress bars corresponding to KCs.}
    \label{fig:apprentice_1}
\end{figure*}

\section{Study 1: Building and Deploying Apprentice Tutors}

Much of existing research on educational technology focuses on deploying intelligent tutors in K–12 settings \cite{koedinger1997intelligent,heffernan2014assistments,aleven2006cognitive}. However, it remains unclear whether these tutors have the same impact on adult learners when used as a supplementary resource. To address this gap and our first research question, we developed an intelligent tutoring system specifically for adult learners, where participation is voluntary. By analyzing tutor adoption and usage over time, this study provides insights into how this understudied user group interacts with educational tools.  

\subsection{Apprentice Tutors}
We developed a web-based intelligent tutoring system platform, called Apprentice Tutors, that is accessible via most modern browsers from a mobile device or a computer, see  Figure~\ref{fig:apprentice_1}. Each tutor supports tutoring of multiple types of problems.
Each problem type consists of an interface to scaffold problem solving and an expert model that can provide feedback on student input.
The Apprentice Tutors architecture was built with Python and the expert models were built with a rete-based production rule engine \cite{forgy1989rete}.

To develop tutors on this platform, we performed a cognitive task analysis with instructors \cite{lovett1998cognitive}. Based on the information from this analysis, we designed and programmed a tutor interface for each specific problem type. Each step within the tutor interface is mapped to a single skill, which is represented in the expert model via a particular production rule.

Each tutor transaction is stored within the Apprentice Tutor platform's database. This allows it to track the progression of skills over time as users continue to interact. We created an adaptive problem selection feature to personalize the learning experience \cite{vanlehn2006}. Depending on how they would like to practice, learners can choose between adaptive problem selection or manually select a particular problem type within each tutor. Each tutor also generates randomized problems for students to practice based on the user interface constraints.

As users interact with dynamically generated problems, the tutors provide real-time correctness feedback and multi-layer hints in a text box below the tutor interface. The Apprentice Tutors platform also has visualization dashboards that allow both students and teachers to track skill mastery \cite{apprentice_dashboard}. Apprentice Tutors allows for the seamless deployment of tutors to classrooms. To facilitate this goal, the interoperability standards for learning tools developed by the IMS Global Learning Consortium \cite{lynch2004interoperability} were used. All major learning management systems comply with these standards and allow educational technologies to be embedded in popular learning management systems, such as Blackboard and Canvas.

\subsubsection{Apprentice Tutors Topics}
For this study, we developed four tutors covering radicals, exponents, factoring polynomials, and rational equations (Figure \ref{fig:apprentice_1}). Aligned with the OpenStax college algebra textbook \cite{openstax_intermediate_algebra_2e}, the tutors support textbook topics. Problem sets were designed to match worked examples, such as creating individual tutor problems for exponent rules like power, product, and quotient.

\subsubsection{Building Expert Models}
A tutoring system incorporating a rule-based expert model is known as a cognitive tutor \cite{anderson_1995}. The expert model generates multiple solutions, enabling personalized, adaptive feedback by evaluating student performance against domain-expert strategies. Combined with a knowledge tracing algorithm \cite{corbett1994knowledge}, it also supports adaptive problem selection. Building expert models for complex tutors is challenging due to step interdependencies and the need for problem-specific and generalized reasoning \cite{anderson_1995}. In Apprentice Tutors, each production rule represents a skill required to complete a step. The production engine, based on \cite{Doorenbos1995ProductionMF}, dynamically composes these rules at runtime to handle varied inputs. Cognitive tutors improve learning by providing immediate, personalized feedback, and adaptive problem sequencing, making them effective for individualized instruction \cite{anderson_1995, vanlehn2006, koedinger1997intelligent}. Each type of problem in Apprentice Tutors has a dedicated interface, expert model, and problem generator. The expert model enables correctness feedback and context-sensitive hints. While developing expert models is time-intensive, it ensures tutors function across randomly generated problems, enhancing adaptability. Grounded in cognitive theory and intelligent tutoring research, this design balances theoretical soundness with practical effectiveness.

\subsubsection{Real-Time Correctness Feedback}
Real-time feedback has been shown to be an effective instructional strategy, particularly in improving learning outcomes and fostering immediate error correction \cite{shute2008focus}. By providing learners with timely feedback, the tutor helps them identify and correct errors before they become ingrained, promoting the mastery of the targeted skills. With the Apprentice Tutors, each rule is designed to correspond to a specific step in the interface, with computations performed by the rules stored as variables. If the student's input matches the output of the specific production function linked to the interface, the input is marked as correct. In this case, the corresponding field is highlighted in green and is disabled to prevent further editing, prompting the user to proceed to the next step. Conversely, if the student's input does not match the output of any production function, it is marked as incorrect, and the field is highlighted in red, prompting the user to retry the step. This functionality enables the tutor to provide real-time correctness feedback to users (as shown in Figure \ref{fig:apprentice_1}, part B). 

\subsubsection{Hinting Strategy}
Each tutor has a dedicated hint box at the bottom of the problem interface, which provides users with the ability to request help. Multiple layers of hints were implemented for each problem type, allowing users to access up to three tiers of assistance (as seen in Figure \ref{fig:apprentice_1}, section D). In the initial hint, the specific step under consideration is highlighted and the user is prompted with a message encouraging them to tackle the particular step. At the second hint level, an explanatory message is delivered in the hint box, offering guidance on how to approach solving the step. In the final hint tier, the solution to the step is revealed, effectively providing the learner with a worked example. This hinting strategy is inspired by \citet{anderson_1995}.

\subsubsection{Adaptive Problem Selection}
The Apprentice Tutors platform provides a personalized tutoring experience through the adaptive problem selection option (as seen in Figure \ref{fig:apprentice_1}, Section A). While each Apprentice tutor has this adaptive selection tab, this is optional, and learners may opt to click on a problem type directly. The adaptive problem selection option personalizes the student's learning experience by providing them with problems that exercise skills they have not yet mastered. Actions such as entering a tutor step or requesting a hint are logged within the tutoring database. Bayesian Knowledge Tracing (BKT) \cite{corbett1994knowledge} is used to assess the student's mastery of each skill. BKT is an approach that models the student's learning using a hidden Markov model. It is commonly used in intelligent tutoring systems \cite{Yudelson2013}. 

\subsubsection{Performance Visualization Dashboards}
Visualizations of student trajectories are increasingly used to comprehend student activity and their navigation through course content \cite{goulden2019ccvis, lundqvist2019visualising}. To facilitate the visualization of skill progression, we developed a dashboard for students to view their mastery on each skill. This dashboard displays the mastery level of each KC, broken down by the specific problems in which these KCs are encountered (as seen in Figure \ref{fig:apprentice_1}, part E). In addition, Guo et al. \cite{apprentice_dashboard} developed visual analytics dashboard for instructors that shows detailed provenance data across multiple coordinated views \cite{apprentice_dashboard}. 

\subsection{Methods}
We developed tutors for four college algebra topics available at several state technical college classes spanning from September 2021 to June 2022. The institution, which focuses on technical 2-year degree programs, primarily serves non-traditional adult learners returning to school to acquire new skills in support of their career goals.  Adult learners present unique challenges and opportunities due to their distinct motivations, cognitive developmental differences, and the voluntary nature of their involvement with educational technology. This study examines how sociotechnical factors influence tutor adoption and usage in self-directed learning contexts, focusing specifically on adult learners. By analyzing their lived experiences and unique needs, we aim to improve the design of intelligent tutors to better support adult learner's educational goals and skill acquisition. The ultimate goal is to enhance learning outcomes for adults, enabling them to succeed in both personal and professional development.

During the deployment, students accessed the tutors through the Blackboard Learn system. The use of the tutors was entirely voluntary, and no course credit or payment was offered for their use. Before and during the deployment, we engaged teachers through email communications and presentations at their monthly department meetings, providing information about the tutors, their features, and potential benefits for students. Before starting this study, its protocol was reviewed and approved by our institutional review board. Upon first accessing a tutor, students were asked to consent to participate in the study. Those who did not consent could not use the tutors. For consenting students, data was collected on usage, including access times, tutor interactions, and student-course correlations.

\begin{figure*}
    \centering
    \includegraphics [width=1\textwidth]{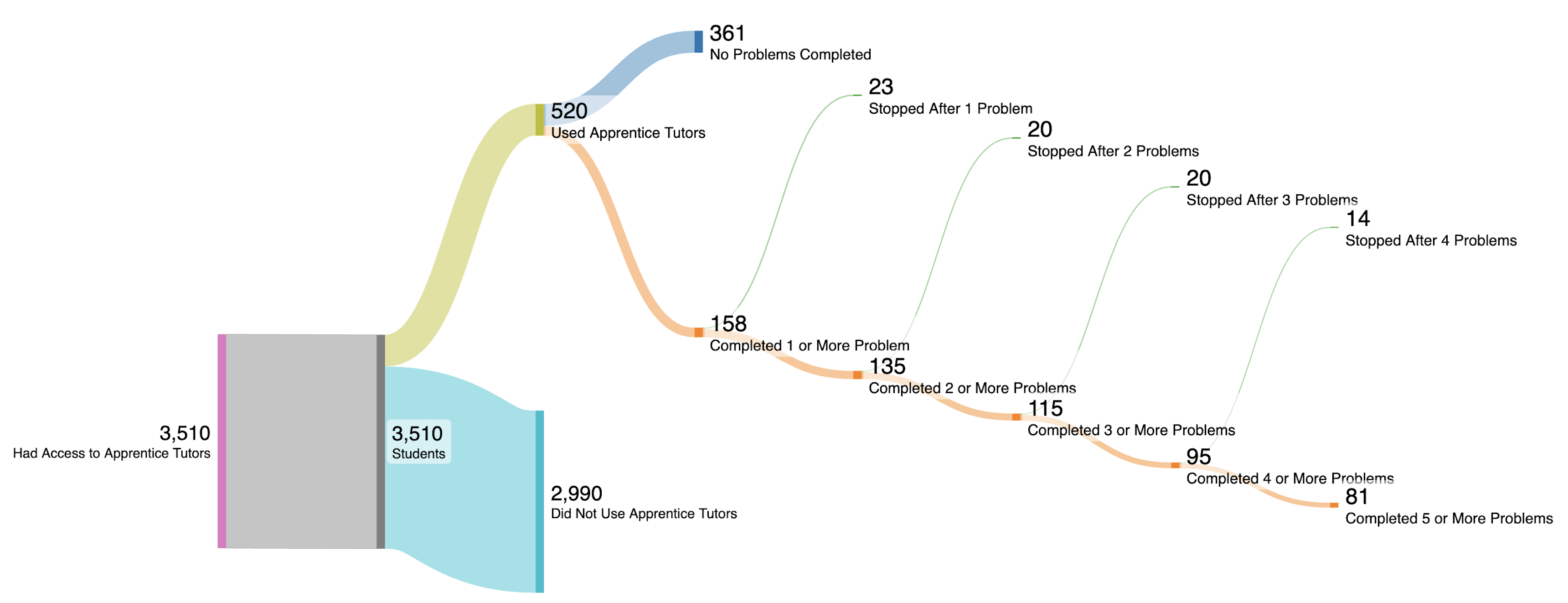} 
    \caption{Graphical representation of student engagement: This figure shows the flow of student interaction with the tutoring program, starting with total access and dividing into students who used or did not use tutors. It tracks progress through problem completion levels, highlighting those who completed no problems, one problem, and multiple problems, with further breakdowns for students stopping at specific milestones (e.g., 2, 3, 4, or 5 problems).}
    \label{fig:funnel}
\end{figure*}

\subsection{Results} 
\subsubsection{Usage and Adoption}
During the deployment period, 3,510 students were granted access to the Apprentice Tutors. Figure \ref{fig:funnel} illustrates the usage and retention funnel throughout the first four academic quarters of deployment. Adoption was defined as the percentage of students who clicked on the tutors at least once, resulting in an overall adoption rate of approximately 14.8\%. In particular, as the number of classes that use the system increased from 4 to 58, the rate of initial clicks decreased from 62.84\% to 13.78\%. A term-by-term breakdown of tutor deployment and adoption metrics are stated in Table \ref{tab:tutor_deployment}.

To evaluate tutor adoption, we assessed the percentage of students who accessed the tutors and solved more than one problem. This metric decreased slightly over time, from 17\% in the first term to 14\% in the fourth term. The participation was highest during the second term, when 22\% of the students who initially accessed the tutors proceeded to solve more than one problem.

\begin{table}[htbp]
\centering
\scriptsize 
\setlength{\tabcolsep}{2pt} 
\renewcommand{\arraystretch}{1.2} 
\begin{tabular}{|p{0.6cm}|p{1.2cm}|p{1cm}|p{1cm}|p{4cm}|}
\hline
Cycle & Term & Start Date & End Date & Metrics \\
\hline
1 & Spring 2022 & 1/3/22 & 5/6/22 & \begin{tabular}[c]{@{}l@{}}Classes Deployed: 4 \\ Students with Access: 76 \\ Students with Interaction: 47 \\ \% of Possible Students Used Tutor: 61.84\% \\ Students Finished $\geq$ 1 Problem: 8 \\ \% of Tutor Users Finished $\geq$ 1 Problem: 17.02\%\end{tabular} \\
\hline
2 & Summer 2022 & 5/9/22 & 8/5/22 & \begin{tabular}[c]{@{}l@{}}Classes Deployed: 31 \\ Students with Access: 265 \\ Students with Interaction: 71 \\ \% of Possible Students Used Tutor: 26.79\% \\ Students Finished $\geq$ 1 Problem: 16 \\ \% of Tutor Users Finished $\geq$ 1 Problem: 22.54\%\end{tabular} \\
\hline
3 & Fall 2022 & 8/15/22 & 12/18/22 & \begin{tabular}[c]{@{}l@{}}Classes Deployed: 64 \\ Students with Access: 2054 \\ Students with Interaction: 229 \\ \% of Possible Students Used Tutor: 11.15\% \\ Students Finished $\geq$ 1 Problem: 30 \\ \% of Tutor Users Finished $\geq$ 1 Problem: 13.10\%\end{tabular} \\
\hline
4 & Spring 2023 & 1/2/23 & 5/14/23 & \begin{tabular}[c]{@{}l@{}}Classes Deployed: 58 \\ Students with Access: 1364 \\ Students with Interaction: 188 \\ \% of Possible Students Used Tutor: 13.78\% \\ Students Finished $\geq$ 1 Problem: 25 \\ \% of Tutor Users Finished $\geq$ 1 Problem: 13.30\%\end{tabular} \\
\hline
\end{tabular}
\caption{Apprentice Tutor Deployment Usage Data by Cycle}
\label{tab:tutor_deployment}
\end{table}

Of the 3,510 students with access, 520 (14.8\%) clicked on the tutors at least once (Figure \ref{fig:funnel}). Within this subgroup, 37 students (0.105\% of the total of 3,510) clicked on the tutors but did not sign the consent form; Users who did not give us consent did not use the Apprentice Tutors. Among these 520 students, 361 did not complete any problems. User interactions were tracked through button clicks, hint requests, and keyboard inputs. Of the 520 who accessed the tutors, 158 (30.4\% of 520) completed at least one problem. A problem was considered complete if all steps were answered correctly and the ``done'' button was clicked. Moreover, 81 of these 158 students (51.3\% of 158) eventually solved five or more problems using Apprentice Tutors.

\subsection{Discussion}
Throughout the first year of deployment, we observed a decrease in overall usage and adoption of tutors as the implementation scale increased from 4 to 58 classes in each term of the academic year (presented in Table \ref{tab:tutor_deployment}. The rate of initial tutor clicks decreased from 62.84\% in the first term to 13.78\% in the fourth term. This downward trend highlights several potential challenges in scaling the deployment of intelligent tutoring systems.

One possible explanation for this finding is the reduced ability to interact directly with the instructors as the number of participating classes increased. At the start of the deployment, we worked closely with the initial group of instructors, providing detailed communication, personalized training sessions, and opportunities to address their concerns. For example, we hosted introduction sessions to explain how the tutors worked, with the goal of making the instructors more comfortable incorporating technology into their classrooms. However, as the scale grew, this level of individual participation became less feasible, potentially affecting the instructor's confidence and willingness to promote the tutors to their students.

Another factor to consider is the varying alignment between the tutors and the instructional styles of different teachers. The tutors were initially co-designed with the input of a small group of instructors, whose preferences and teaching methods shaped the features and content of the system. As the deployment expanded, teachers who were not involved in the design process may have found the tutors less intuitive or less aligned with their pedagogical practices. This misalignment could have reduced their enthusiasm for integrating technology into their lessons and encouraging students to use it.

The voluntary nature of tutor usage may have contributed to the decline in adoption rates. Instructors and students alike can prioritize tools that are directly tied to grades or other extrinsic motivators. Without a strong incentive to explore and adopt the tutors, some students and teachers may have opted to focus on other resources perceived as more extrinsically beneficial. These findings suggest the importance of understanding and addressing the sociotechnical factors that influence adoption. Instructor participation, alignment with teaching practices, and the presence of motivating factors for voluntary use are critical to promoting the adoption and sustained use of intelligent tutoring systems among adult learners.

To address this challenge, we propose two potential strategies for future large-scale tutor deployments targeting adult learners. First, a robust tutorial that clearly demonstrates the benefits and functionality of the intelligent tutor could help mitigate barriers to understanding, especially when individual face-to-face instructor engagement is not feasible. Second, automating engagement with both instructors and adult learners is recommended as an area for further research, potentially through personalized communication strategies that speak to the motivations and needs of adult learners.

Despite the lower adoption rates, we saw encouraging signs of user retention in tutors. Among the 3,510 students granted access, 520 (14.8\%) actively clicked on the tutors. Of these 520 adopters, 158 (30.4\%) solved at least one problem, and 81 (51.3\% of the 158) proceeded to solve five or more problems. An additional term-by-term analysis revealed that the percentage of students who solved more than one problem shifted from 17\% to 14\% between the first and fourth quarters, reaching a peak at 22\% during the second term. Although these findings provide preliminary evidence that tutors support learning and retention of usage among adult users, more research is needed to rigorously measure the causal effects of these systems on adult learning outcomes. 

\section{Study 2: Investigation of Challenges}
The deployment of Apprentice Tutors showed a positive trend in user retention, where nearly half of the users who solved at least one problem continued to solve at least five or more unique problems. Despite this, the overall adoption rate of these tutoring systems remained at 14.8\%. Our analyses suggest an inverse relationship between the deployment scale and adoption rates. A plausible reason for the lower adoption of tutors over time may have been the decrease in direct teacher support as the tutor deployment scaled. This support included an email nudge approach to remind teachers to further remind their students that tutors were available as a resource to use.  

However, we were interested in investigating additional factors that drive adoption among adult learners. To identify these factors, we conducted a focus group study that involved collecting and analyzing feedback from students and teachers who engaged with Apprentice Tutors during the academic year. Analyzing this user feedback helped to surface several critical themes, providing a foundational basis for developers and educational content designers to better understand the needs and challenges of adult learners. 

\subsection{Methods}
\subsubsection{Recruitment}
To investigate our research questions, we conducted five semi-structured focus group sessions with both students (who used the tutors) and teachers (who agreed to tutors being deployed in their classrooms). We recruited participants through snowball sampling, through direct e-email, and newsletter postings. Throughout the five sessions, we collected data from fifty-three participants, which consisted of forty-two students and eleven teachers. Participation in these focus groups was optional and a 15 dollar gift card was available to participants after the completion of the session. For the purposes of these focus groups, age and demographics were not used in the filtering or selection of the participants. All participants were located in the United States and all focus groups were conducted virtually. All virtual focus groups were about 60 minutes.

\subsubsection{Teacher Engagement}
Throughout the deployment, we actively engaged with teachers via email to support the adoption of the Apprentice Tutors. During the first two terms, we offered information sessions and in-person tutorials to help teachers become more comfortable with using the tutors. However, in the third and fourth terms, the scale of the deployment made individualized assistance impractical. Instead, we worked closely with a key point of contact who facilitated the deployment by sharing updates, tutorials, and other relevant information with the teachers.

\subsubsection{Data Collection}
All participants provided their written consent to participate in the focus group prior to beginning the sessions. The discussions were grounded in key experiential feedback on how participants responded to the tutoring systems. The discussions were led by moderators who guided participants through a series of open-ended questions stated below. 

\begin{itemize}
    \item Describe the main benefits of tutors?
    \item Describe any frustrations in using the tutors?
    \item Were the hints provided adequate?
    \item Were you able to easily understand how to use the tutors?
    \item Describe your comfort in recommending the tutors?
    \item Did the tutors reflect the course material correctly?
    \item Is there anything else you would like to add to the tutors?
\end{itemize} 

The prompts presented were open-ended and semi-flexible, where participants were given the opportunity to reflect more broadly on their experiences through other academic contexts.

\begin{table*}[p]
\centering
\begin{tabular}{|c|p{7cm}|p{9cm}|}
\hline
\textbf{} & \textbf{Key Intelligent Tutor Themes} & \textbf{Participant Quotes from Focus Groups} \\
\hline
\multirow{3}{*}{\rotatebox[origin=c]{90}{AI support Feelings  }} & Users are concerned that tutors will decrease the overall amount of ``human'' and ``human-like'' interaction, which they think will decrease learning & S: {\it ``I think one thing that's true of AI, whether that's a tutor or not, is it takes out the human factor. The AI isn't gonna understand that there might be some barriers from it trying to walk you through something that it can't see.''}\\
\cline{2-3}
& Users like ``superhuman'' aspects of tutor, where it provides support a human teacher cannot \newline & T: {\it ``The positive aspect for me is knowing that there's something out there to help strengthen those concepts that students have seen before but haven't seen in a while and they may need some more guided, some more practice on them or I can't, I can't spend a lot of time on a particular topic but maybe they can get more practice on that topic if they're struggling on that topic.''} \\
\hline
\multirow{3}{*}{\rotatebox[origin=c]{90}{Tutor Hints Support  }} & Users find current hints and feedback to be insufficient and want clearer, more detailed explanations that help students understand what they did wrong, how the tutor got the answers it got, and that help them get ``unstuck'' & S: {\it ``So I click next hint and it just gives me the answer and I don't know how it got that answer so that it didn't click for me. I was just, I kind of gave up on using it.''} \\
\cline{2-3}
& Users want tutors to explicitly connect and link to relevant concepts and explanatory content & T: {\it ``When I finally went into the OpenStax textbook and I'm like, OK, that's what they're referring to as B and C and I'm like, they're defined as this and this, I think that would be helpful is if that were clearly displayed, you know. An example of, here's an example of a problem this is what we think.''} \\
\cline{2-3}
& Users would like tutors to link to explanatory videos to support their learning & S: {\it ``YouTube as well [for getting unstuck on a problem]. There's, a lot of teachers on there and sometimes they just explain it different and I get it from watching several different peoples' method of doing it.''} \\
\cline{2-3}
& Users think it would be helpful if they could ask the tutor questions and have dialogue-based interactions & S: [What would be a characteristic of an ideal tutor?] {\it ``Something that's gonna have a conversation with you rather than feeling like I'm just plugging something into a calculator.''} \\
\hline
\multirow{3}{*}{\rotatebox[origin=c]{90}{\small Tool Adoption  }} & Teachers want more tutor content and they want to be able to create and customize it & T: {\it ``I don't know if I have the skills to build my own tutor, but it would nice to create my own tutor problems''} \\
\cline{2-3}
& Teacher adoption depends on them understanding how tutors work and seeing alignment between tutors and course content & T: {\it ``If I am confused on what the tutor interface says, how can I tell the students to use the tutors?''} \\
\cline{2-3}
& Users are willing to provide time and data to improve the tutor & {\it T: Two teachers - In response to the question of whether they would be willing to collaborate with the development team to address the deficiencies in the tutor: ''Yes, absolutely'' } \\
\hline
\multirow{3}{*}{\rotatebox[origin=c]{90}{Usability \& Value Considerations  }} & Users are more likely to adopt tutors if they see a clear benefit/incentive and if they are ``reminded'' & {\it T: ``I don't think I have done a good job reminding students to do the tutor''} \\
\cline{2-3}
&  Users find tutors frustrating \& confusing and we need to increase their usability and provide more support (e.g., tutorial videos) before their value will be realized & T: {\it ``Now the concept is great... You know I don't delve deeply into simplifying radicals because I don't have the time. But if it was one that was easy to understand and easy to get through and user friendly, Amen.''} \\
\cline{2-3}
& Users had many usability/bug issues that produced confusion & {\it S: ``Let's pull up the exponents product rule, there was no way to show what we need to type in the first box? In this particular problem, what is the correct answer here [the hint box showed the answer in LaTeX notation which was hard to understand]''} \\
\cline{2-3}
& Teachers are frustrated they cannot see who is using the tutor, how tutor use relates to student progress/learning, or evidence of tutor effectiveness. & T: {\it ``I was also frustrated, I couldn't see who was accessing it. I would have to rely on students [to tell me]''} \\
\cline{2-3}
& Teachers found simpler tutors less confusing & T: {\it ``The tutor is pretty, like it kind of speaks for itself since it's pretty simplistic.''} \\
\hline
\multirow{3}{*}{\rotatebox[origin=c]{90}{Features  }} & Users had several features they liked also suggested new possible features & S: {\it ``Is there a possibility for if you get the question incorrect after so many times, it tells you how to move on.''} \\
\cline{2-3}
& There are a couple of key subpopulations (minors and neurodiverse user demographics) we should identify and design for & S: {\it ``I have a learning disability called Dyscalculia. And the unfortunate bit is the AI tutor doesn't seem to take that into consideration.''} \\
\hline
\end{tabular}
\caption{Thematic analysis of stakeholder perspectives on Apprentice tutors: This table synthesizes feedback from students (S) and teachers (T) regarding their experiences with AI tutoring systems.}
\label{tab:affinity}
\end{table*}

\subsubsection{Data Analysis}
Video recordings and verbal transcriptions of focus groups were stored to capture data for analysis. To perform our analysis, we used the Braun and Clarke reflexive thematic analysis framework \cite{Braun2006, Braun2019, Braun2021}. This method, grounded in a post-positivist paradigm, emphasizes the researcher's active role in knowledge creation and takes into account the philosophical and theoretical underpinnings of the analytical process \cite{Braun2019}. Reflexive thematic analysis provides greater flexibility compared to approaches that require codebooks and reliability calculations between players, allowing dynamic engagement with the data while maintaining analytical rigor \cite{Braun2019}. In addition, this approach promotes the construction of collaborative themes through ongoing dialogue among researchers.

The data analysis process was conducted virtually over a three-week period by six researchers. Four 90-minute collaborative sessions were held, following the analysis procedures outlined by Braun and Clarke \cite{Braun2006}. Each author coded approximately eight to nine transcripts, collectively producing several hundred codes. Throughout each session, continuous discussions about codes, themes, and key findings were held.

For organizing and categorizing codes, Miro (an online virtual whiteboard tool) was used to facilitate collaborative analysis. Subsequent sessions featured in-depth reviews and discussions of the codes to form themes from the coded data. This open-ended approach allowed exploration without relying on predefined categories. Table \ref{tab:affinity} presents the resulting taxonomy, which outlines overall themes, sub-themes, and examples of coded data. Although it offers a simplified view of the themes and quotes supporting the core recommendations, the complete analysis yielded 17 recommendations and 48 themes derived from several hundred coded excerpts.

This methodological approach allowed for a thorough investigation of adult learners’ experiences and needs in relation to AI-powered tutoring systems, ensuring a complete understanding of the focus group data.

\subsection{Results}
We identified key key adult learners' needs that AI-powered tutoring systems could address through analyzing the focus group data. To do so, we conducted a bottom-up analytical approach within Miro, where raw data was mapped onto color-coded digital post-it notes. Through iterative analysis, we identified several emergent themes, which are summarized in Table \ref{tab:affinity}. This table presents the main theme categories, specific themes, and exemplar user needs within each theme.

The analysis yielded five primary categories of insights: perceptions of AI support, need for enhanced explanatory capabilities, pathways for improved tool adoption, usability and value considerations, and suggestions for additional features. Each category encapsulates the concerns, preferences, and recommendations of various stakeholders, highlighting areas for potential enhancement in AI tutoring technologies to better serve adult learners and their educators.

A prominent theme that emerged was the perceived value of AI tutors' ``superhuman'' capabilities. Participants appreciated the system's ability to provide support in situations where human teachers might be unavailable or less effective. One student emphasized the importance of flexibility:  {\it ``The flexibility is key—balancing school, work, and other commitments, the ability to access the tutor anytime, anywhere, like during a lunch break, is incredibly beneficial.''} This sentiment was echoed by an educator who noted: {\it``Knowing that students have access to help anytime removes barriers to learning. It addresses common excuses, such as not finding help late at night, by providing a resource like the exponent tutors available at all hours.''} These remarks support the initial hypothesis regarding the suitability of AI tutors for adult learners, highlighting the value placed on flexible, round-the-clock support.

We received recurring feedback on tutor requirements through focus groups with teachers and students. The data suggested that adoption might improve if the tutors were: (a) easy to understand, (b) easy to navigate, and (c) user-friendly. One teacher described this perspective as follows:
\emph{``Now the concept is great. What I was thinking when I first heard of this is, you know I don't delve deeply into simplifying radicals because I don't have the time. If I had somewhere where they could go practice on simplifying radicals, that would be great. But if it was one that was easy to understand and easy to get through and user friendly, Amen.''}

As shown in Table \ref{tab:affinity}, five key themes emerged from these focus groups:

\begin{itemize}
\item \textbf{AI Support Perceptions}: Concerns were raised that tutors could reduce the total amount of ``human'' or ``human-like'' interaction (\emph{``It takes out the human factor... the AI isn't gonna understand that there might be some barriers...''}), but users also valued ``superhuman'' capabilities beyond what a single instructor can provide (\emph{``If they can get more practice on that topic... maybe they can get more practice on that topic if they're struggling...''}).

\item \textbf{Tutor Hints Support}: Participants indicated that current hints and feedback are insufficient and need to be more detailed, including clear explanations of how answers are obtained (\emph{``I click next hint and it just gives me the answer and I don't know how it got that answer...''}). Requests were also made for links to relevant content, explanatory videos, and the ability to ask the tutor direct questions.

\item \textbf{Tools Adoption}: Teachers expressed a need for more tutor content and customizable options (\emph{``You guys need to make this similar to a homework platform...''}). Adoption also depended on understanding how tutors work, seeing alignment between tutors and the curriculum, and having sufficient time to integrate tutors into the classroom.

\item \textbf{Usability \& Value Considerations}: The participants wanted a clearer incentive or benefit for using the tutors and emphasized the importance of reminders. Some found the tutors confusing or difficult to use, highlighting a need for increased usability, tutorials, and reduced bugs (\emph{``When you go to the radicals tutor, there were so many unnecessary steps...''}). Teachers also wanted visibility into how students used tutors and how usage was related to learning outcomes.

\item \textbf{More Features}: Several preferred features were mentioned, such as greater flexibility in problem-solving methods (\emph{``There's also a bunch of different options you can choose for the problem...''}) and specialized support for subpopulations (e.g., minors, neurodiverse learners) with specific needs (\emph{``I have a learning disability called Dyscalculia. And the unfortunate bit is the AI tutor doesn't seem to take that into consideration.''}).
\end{itemize}

Based on these themes, we derived the following recommendations for tutoring platforms beyond Apprentice Tutors.

\begin{itemize}
    \item \textbf{Incorporate human elements}: To address concerns about reduced human interaction, features that simulate human-like dialogue (e.g., ability to send teachers a note) are recommended. These features respond to feedback indicating that users value personalized, interactive support that reduces feelings of isolation and facilitates deeper engagement. Research on social presence in online learning \cite{richardson2017social} supports the importance of such features in fostering a sense of connection.

    \item \textbf{Align with curriculum}: Co-designing intelligent tutor content to integrate seamlessly with existing course materials is crucial. While the content was aligned with the curriculum, teachers often struggled to see the explicit mapping between the tutor content and their instructional goals, leading to frustration and the perception of tutors as ``extra work.'' Instructional alignment, when clearly communicated and supported, is a well-established factor in improving learning outcomes \cite{anderson2001taxonomy}. 
    
    \item \textbf{Demonstrate value clearly}: Tutoring platforms should transparently showcase their benefits, such as highlighting ``superhuman'' capabilities (e.g., providing instant feedback or adaptive problem selection) and offering visible progress indicators for students. Focus group participants emphasized the importance of clear, visible benefits to motivate sustained use and build trust in the platform's effectiveness. Additionally, incorporating visualization dashboards for both teachers and students can enhance the user experience. For teachers, dashboards can provide insights into student progress, performance trends, and areas needing additional support. For students, progress dashboards can serve as motivational tools, allowing them to track their achievements and understand their learning journey.
    
    \item \textbf{Streamline onboarding}: An intuitive onboarding experience minimizes the initial cognitive load for both students and teachers, addressing common feedback on frustration and confusion when interacting with the platform. Effective onboarding may include tutorial videos, step-by-step guides, and streamlined user interfaces, which align with research on reducing extraneous cognitive load to improve usability \cite{sweller1988cognitive}.

    \item \textbf{Harmonize teaching approaches}: Positioning the tutor as an extension of classroom instruction, rather than a replacement, is essential to facilitate greater adoption. While the system was positioned in this way, some teachers still perceived it as potentially competing with their role. To address this, framing the technology as ``AI-guided practice'' rather than ``tutors'' may better communicate its purpose as a tool to reinforce learning rather than teach independently. This could ease concerns about dehumanization, clarify its intended use, and empower educators to integrate it more effectively into their course.   
\end{itemize}

These recommendations are designed to address the needs and concerns identified in focus groups, potentially enhancing the effectiveness and adoption of AI tutoring systems for adult learners.

\subsection{Discussion}
The focus group sessions and subsequent data analysis revealed a complex interplay of factors that can hinder the wider adoption of tutoring systems among adult learners. Although the results section highlighted several positive aspects and demonstrated the potential of these systems, a number of additional insights surfaced that point to challenges in real-world implementation.

A key theme involved the value of tutors for non-traditional students returning to education after extended gaps. As noted in the results section, participants described these technologies as especially useful for refreshing foundational skills, suggesting that tutors may help foster academic re-engagement among adults who have been distant from formal study for significant periods. However, instances of user frustration were also identified, primarily resulting from difficulties in navigating tutor interfaces and integrating tutor activities into broader learning tasks. The lack of tutorials or instructional videos was frequently cited as a barrier, reinforcing the idea that onboarding processes should be intuitive and transparent.

Another issue concerned the diversity of teaching methodologies within the same academic subject. Disparities between tutors’ instructional methods and the approaches used by some instructors led to confusion and, ultimately, reduced buy-in. As noted in the results, tutor customization emerged as a recurrent topic: teachers expressed a desire for flexible authoring tools that would let them customize tutors to more closely align with their unique pedagogical styles. The need for instructor involvement was further emphasized by the limited incentives that sometimes restricted the willingness of teachers to invest time learning new technologies. In addition, the tutor system examined in this study was limited to a small set of course chapters, and many educators wanted concrete, visual evidence of tutor effectiveness - preferences that, if not met, could diminish adoption rates.

We found that sociotechnical factors played an important role in the adoption of tutors beyond just effectiveness. These factors include ease of use, technological infrastructure, scheduling constraints, and alignment with curriculum objectives. Adult learners, in particular, can face additional barriers such as limited time, competing responsibilities, and varying levels of digital fluency. These elements underscore the importance of holistic approaches to designing and deploying AI tutors that address not only instructional goals but also practical considerations of user contexts.

Subsequently, the recommendations we provided are grounded in qualitative analysis of user feedback, which highlighted both recurring challenges and opportunities for improvement. For instance, the emphasis on curriculum alignment stems from teacher's frustration with tutors that were perceived as disconnected from their lesson plans, often resulting in reduced adoption. Similarly, addressing concerns about the impersonal nature of AI systems requires focusing on how the technology can support students in need of human assistance. For example, intelligent tutors can help by reducing the workload on teachers, enabling them to allocate more time and attention to students who require additional support. By providing targeted feedback, these systems can create opportunities for more meaningful teacher-student interactions.

Feedback on the complexity and confusion of tutor interfaces informed the need for streamlined onboarding processes, which aim to reduce cognitive load and improve the initial user experience. By addressing these pain points and integrating the insights from the focus groups, these recommendations aim to create a more user-centric tutoring platform that could improve overall adoption and recurring usage.

\section{Limitation and Future Works}
The Apprentice Tutors were built as a testbed to investigate how to design and deploy tutors for adult learning contexts. Throughout these deployments, several limitations were identified that may have affected overall tutor usage and adoption. A primary concern involves the lack of incentives for teachers to integrate such technologies into their classrooms. Teachers’ busy schedules require tutors that blend seamlessly with existing workflows, without requiring a steep learning curve.

The findings also indicate that teachers expressed their hesitation in part because the system covered only the initial course chapters and lacked a compelling demonstration of tutor effectiveness through visual representations of student performance. To address these issues, future work will focus on redesigning the tutors based on the recommendations provided, incorporating instructional materials such as textbook resources and lecture videos. This approach is intended to enrich the learning experience by offering more comprehensive hints and aligning more closely with the course content.

In addition, future work should expand tutor coverage to encompass a broader range of course material, and model tracing will be integrated alongside knowledge tracing and learning analytics visualizations. These enhancements will provide users with more robust tools to track progress and better understand the problem-solving steps. Future research will involve redeploying these improved tutors and examining their impact over time through longitudinal studies across multiple deployments. Such investigations will facilitate a deeper understanding of how to develop scalable pedagogical technologies that effectively support adult learning.

\section{Conclusion}

There is an unexplored potential in understanding how intelligent tutors can impact adult learners and whether population would adopt educational technologies. The Apprentice Tutors were developed and deployed as a case study to explore the specific needs of adult learners who engage with intelligent tutoring systems. Our preliminary work yielded quantitative tutor usage and user retention after several problems were completed. Despite these positive results, a gradual decline in the adoption of these tutors was observed over time. This observation led to a subsequent study that aimed to elicit teacher and student experiences with and perceptions of intelligent tutors in adult learning contexts. In this second study, we conducted focus groups to collect qualitative insights from both students and teachers who interacted with the tutors. Analysis of this data through thematic analysis revealed several themes that informed a strategic roadmap for enhancing the Apprentice Tutors platform. Furthermore, these insights were synthesized into general recommendations for developers and instructional designers, aimed at improving the engagement and adoption rates of intelligent tutoring systems more broadly. The goal of this research is to contribute to the widespread implementation of intelligent tutors, ensuring that these advanced educational tools are accessible and equitable to adult learners seeking opportunities for educational growth.

\bibliographystyle{ACM-Reference-Format}
\bibliography{main}

\end{document}